\documentclass[pop,fleqn]{w-art}
\usepackage{times}
\usepackage{w-thm}

\usepackage{graphicx}
\usepackage{epstopdf}

\usepackage[dvips]{texdraw}
\usepackage{amssymb}

\newcommand{\AdSxS}{AdS$_5\times{\mathcal S}^5$}

\begin{document}
\DOIsuffix{theDOIsuffix}
\pagespan{3}{}
\Receiveddate{15 November 2003}
\Reviseddate{30 November 2003}
\Accepteddate{2 December 2003}
\Dateposted{3 December 2003}
\title[Chronology Protection in AdS/CFT]{Chronology Protection in AdS/CFT}
\author[F. Author]{Marco M. Caldarelli\inst{1,2}%
  \footnote{Corresponding author\quad E-mail:~\textsf{marco.caldarelli@mi.infn.it}}}
\address[\inst{1}]{Dipartimento di Fisica dell'Universit\`a di Milano, Via Celoria 16, I-20133 Milano.}
\address[\inst{2}]{INFN, Sezione di Milano, Via Celoria 16, I-20133 Milano.}

\maketitle

\begin{abstract}
We review the issue of chronology protection and show how string theory can solve it in the half BPS sector of AdS/CFT. According to the LLM prescription, half BPS excitations of AdS$_5$ $\times$ ${\cal S}^5$ geometries in type IIB string theory can be mapped into free fermion configurations. We show that unitarity of the theory describing these fermions is intimately related to the protection of the chronology in the dual geometries.
\end{abstract}


The principle of causality stands at the foundation of all our understanding of physics. It cannot be relaxed without falling into paradoxes which, in the final analysis, consist either in the attempt to change the past conditions at the origin of the event or in the so-called bootstrap paradox, occurring when the effect is the cause of its own existence. 
The concept of causality is so important that it has been hardcoded into newtonian physics and special relativity. But as soon one takes into account the gravitational force, space and time become dynamical. Their evolution is driven by local equations and the equivalence principle ensures that the causality requirements of special relativity are locally respected. However, the gravitational field equations do not put any global constraint on the topology of time, leaving the possibility for the existence of closed timelike curves (CTCs), seeds of the violation of causality. The global nature of CTCs is what makes them so elusive and difficult to treat.

Essentially, CTCs can arise by two mechanisms, rotation (or dragging of the inertial frames) and effectively superluminal propagation. The most well-known example of non topologically trivial CTCs induced by rotation is given by the G\"odel universe \cite{Godel:1949ga}, a homogeneous and simply connected manifold described by the metric
\begin{equation}
ds^2=-\left(dt+2a\sinh^2\frac{r}{2}\ d\phi\right)^2+dr^2+\sinh^2r\ d\phi^2+dz^2.
\end{equation}
Here, $\phi$ is a periodic angular coordinate, whose orbits become timelike for large $r$ due to the frame-dragging term in the metric and form therefore CTCs. The fact that these CTCs exist at all times is a reason one can invoke to discard this solution. Another example is the Kerr black hole that has harmless CTCs induced by its rotation, but they are hidden by the Cauchy horizon, which is unstable. The situation is worse in higher dimensions, where naked CTCs are a generic feature of Kerr black holes \cite{Myers:1986un}. It should be noted that supersymmetry does not help to rule out chronological pathologies \cite{Gibbons:1999uv}, nor does the lifting to higher dimensional supergravities \cite{Caldarelli:2001iq}.

\begin{figure}[tbp]
\centering
\quad\includegraphics[width=.3\linewidth]{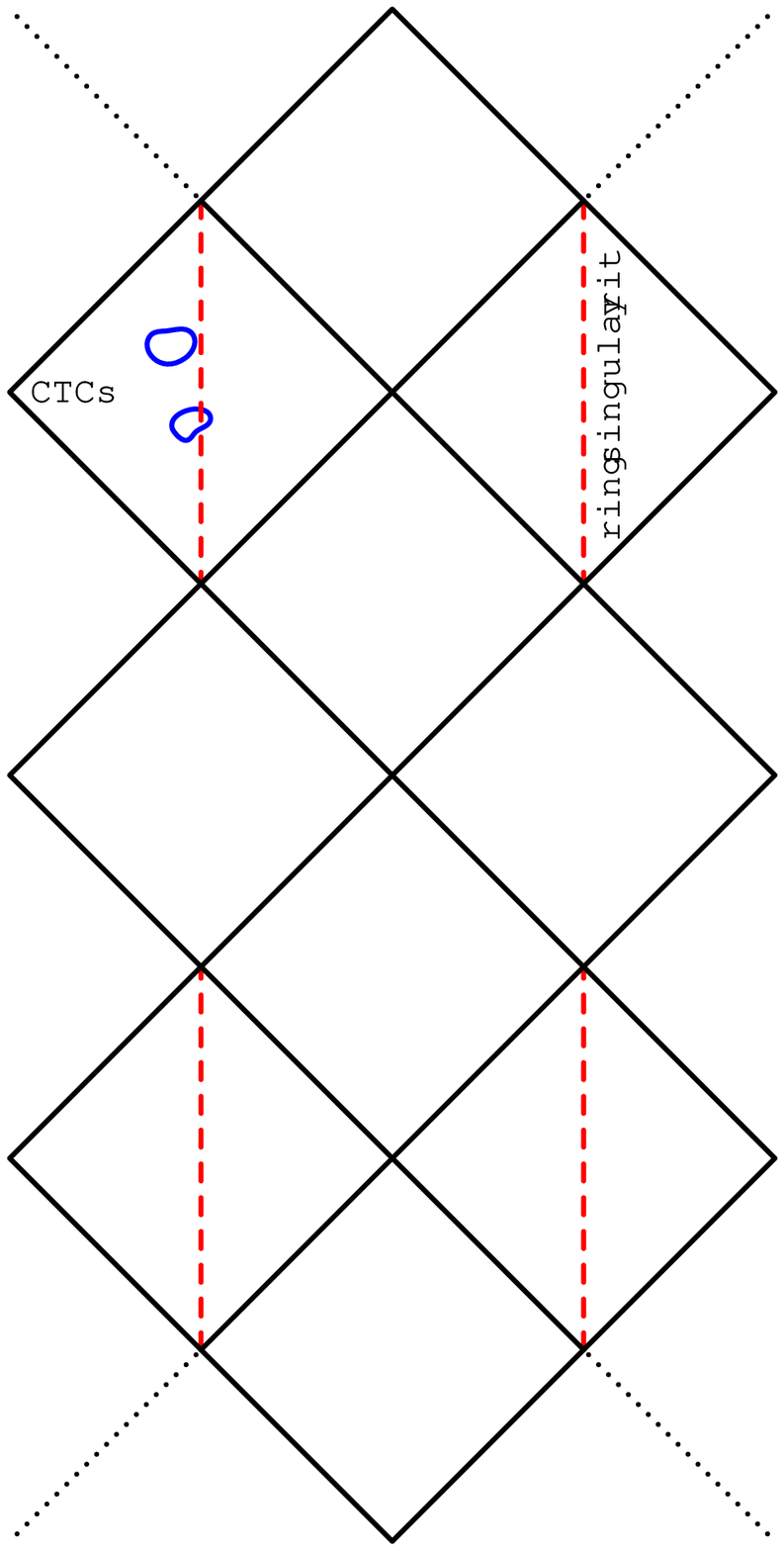}\hfill
\includegraphics[width=.6\linewidth]{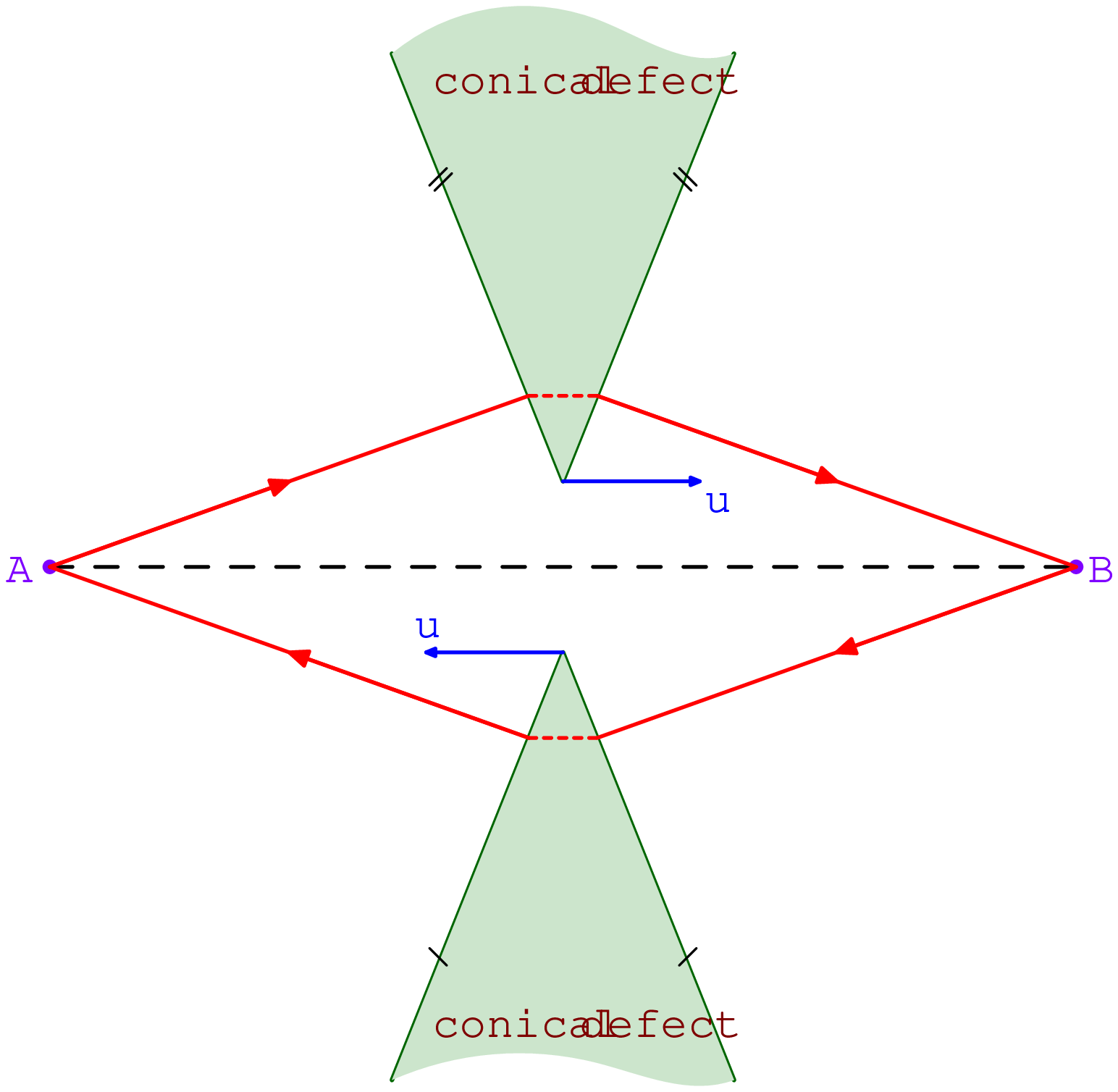}
\caption{On the left, the Carte-Penrose diagram for the Kerr black hole and its CTCs hidden behind the Cauchy horizon. On the right, the Gott time machine and a CTC on this geometry.}
\label{fig-gott}\end{figure}

The other mechanism to create CTCs is to construct geometries allowing for effectively superluminal propagation of signals. For example, Gott's time machine \cite{Gott:1990zr} consists in two parallel cosmic strings in relative motion. The effect of the cosmic strings on the spacetime is to create conical defects, which can be cleverly exploited to build CTCs using the properties of Lorentz transformations (see figure~\ref{fig-gott}). Again, this solution has problems, for example its spacelike total energy momentum, which make it unphysical \cite{Carroll:1994hz}.
However, Ori has recently been able to show the existence of a family of solutions representing time machines with a compact vacuum core \cite{Ori:2005ht}, in which CTCs are created in a compact region evolved from a causally well-behaved past. These solutions satisfy all energy conditions, are regular, asymptotically flat and topologically trivial, proving that general relativity alone cannot prevent the formation of CTCs.

In 1992, Hawking proposed the chronology protection conjecture \cite{Hawking:1991nk}, which states that the laws of physics do not allow the appearance of CTCs. In particular, he argued that the strong backreaction due to quantum fluctuations should destroy the chronology horizon. However, in 1996, Kay, Radzikowski and Wald showed that there are points on the chronology horizon where the semiclassical Einstein's equations fail to hold \cite{Kay:1996hj}, and therefore the ultimate resolution of the chronology protection issue must involve the full theory of quantum gravity (see Matt Visser's review \cite{Visser:2002ua}).
String theory is believed to yield a consistent theory of quantum gravity,
therefore we shall take a more fundamental point of view, and consider the (super-)gravity theories merely as its effective low energy limits. Then, even if the effective theory cannot distinguish between causally well-behaved and ill spacetimes, one can hope that string theory provides a mechanism to rule out the pathological solutions or eliminate the CTCs.

To be more specific, we will concentrate on the half BPS sector of AdS/CFT \cite{Maldacena:1997re}, where we have a strong control on the physics.
Using the AdS/CFT correspondence, we have a non-perturbative formulation of type IIB superstring theory in terms of ${\mathcal N}=4$ SYM in four dimensions \cite{Maldacena:1997re}. In the half BPS subsector, the SYM theory reduces to a gauged quantum mechanical system of $2N^2$ harmonic oscillators, with the harmonic potential generated by the conformal coupling to the $\mathcal{S}^3\times\mathbb{R}$ background. Half of these oscillators consist in left-moving excitations, the other half in right-moving ones. The BPS condition translates then in having just left-moving excitations on-shell. Moreover, the model is gauged, and only gauge-invariant states are physical. One can use the available gauge invariance to diagonalize the matrix of left-movers, thus leaving only $N$ physical degrees of freedom. Performing carefully this change of variables, one introduces a jacobian in the path integral, and effectively changes the statistics of the degrees of freedom which become fermionic.

\begin{figure}[tbp]
\centering
\includegraphics[height=.4\linewidth]{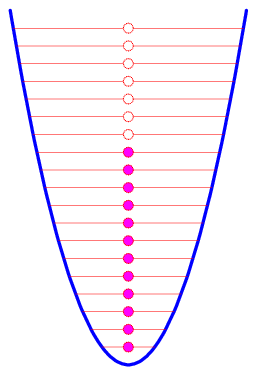}\hfill
\includegraphics[height=.4\linewidth]{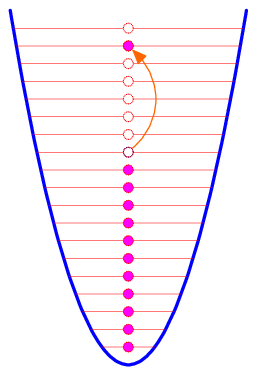}\hfill
\includegraphics[height=.4\linewidth]{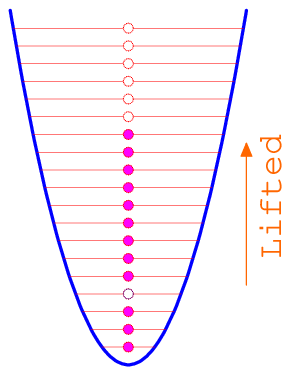}\\
\makebox[.3\linewidth]{a) Fermi sea: ground state}\hfill
\makebox[.3\linewidth]{b) Excitation: giant graviton in AdS$_5$}\hfill
\makebox[.3\linewidth]{c) Hole: giant graviton in $\mathcal{S}^5$}
\caption{Excitations of the Fermi sea (a) have a dual interpretation in terms of D3-brane excitations in the bulk spacetime. The finite depth of the Fermi sea agrees with the stringy exclusion principle.}
\label{fig-harmonic}\end{figure}

In conclusion, the half BPS sector is described by a system of $N$ free fermions in a harmonic well. Its ground state corresponds to the $N$ fermions distibuted on the $N$ lowest energy levels of the harmonic oscillator, while the excitations correspond to excited fermions with energy higher than the Fermi energy, or holes inside the Fermi sea. Using the AdS/CFT correspondence, one can check that the former correspond to giant gravitons growing on the $\mathcal{S}^5$ factor of the metric, and the latter to giant gravitons growing on the AdS$_5$ factor \cite{Berenstein:2004kk,Caldarelli:2004ig} (see figure~\ref{fig-harmonic}).

\begin{figure}[htbp]
\centering
\includegraphics[]{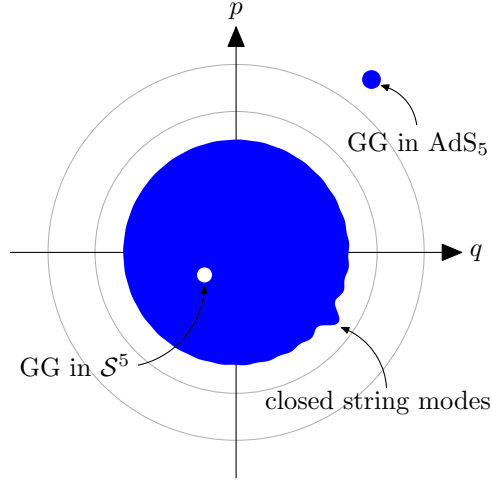}
\caption{Phase space of the harmonic oscillator. Ripples on the Fermi surface correspond to supergravity excitations, droplets above the Fermi sea to giant gravitons in AdS$_5$ and holes in the Fermi sea to giant gravitons in $\mathcal{S}^5$.}\label{fig-phasespace}\end{figure}

To obtain the (semi-)classical configurations of the system, we trade the energy eigenstate basis for coherent states. Then, each fermion occupies an area $\hbar$ in the 
phase space of the harmonic oscillator, and the classical configurations are given by quantum Hall droplets in this space, i.e.~by the description of a region of area $\hbar N$ where the fermions are localized \footnote{The Fermi cell has area $\hbar=2\pi l^4_p$, with $l_p$ the Planck length, and the radius $L$ of AdS$_5$ is related to them by $L^4=4\pi l^4_pN$.}. The ground state of the system corresponds to a disk centered in the origin, of area $N\hbar$. Small deformation of the surface of the disk correspond to closed string excitations, or perturbative gravitons, propagating in the bulk.  Finally, small holes in the disk correspond to stacks of $\mathcal{S}^5$ giant gravitons and droplets of fermions high above the Fermi sea to
AdS$_5$ giant gravitons (see figure~\ref{fig-phasespace}).

Thanks to the work by Lin, Lunin and Maldacena \cite{Lin:2004nb}, we have the complete supergravity solution dual to these classical CFT states. They define a function $z(x^1,x^2,0)$ which takes the value $-1/2$ inside the Hall droplets, and $1/2$ outside. The gravitational dual to this configuration is then given by the metric \footnote{For the five-form field strength we refer to the original paper \cite{Lin:2004nb}.}
\begin{equation}
	ds^2=-h^{-2}\left(dt+V_idx^i\right)^2+h^2\left(dy^2+dx^idx^i\right)
		+ye^Gd\Omega_3^2+ye^{-G}d\tilde\Omega_3^2\,,
\label{LLM}\end{equation}
with
\begin{equation}
	h^{-2}=\frac{2y}{\sqrt{1-4z^2}}\,,\qquad e^{2G}=\frac{1+2z}{1-2z}\,.
\end{equation}
The function $z(x^i,y)$ and the shift vector $V_i$ are determined by fermion configurations $z(x^i,0)$, which act as boundary value,
\begin{equation}
	z(x^1, x^2, y)=\frac{y^2}\pi\int\frac{z(x'_1,x'_2,0)\,dx'_1\,dx'_2}{\left[
	\left({\bf x}-{\bf x}'\right)^2+y^2\right]^2}\,,
\end{equation}
\begin{equation}
V_i(x^1,x^2,y)=\frac{\epsilon_{ij}}\pi\int\frac{z(x'_1,x'_2,0)(x_j-x_j')\,dx'_1\,dx'_2}{\left[\left({\bf x}-{\bf x}'\right)^2+y^2\right]^2}\,.
\end{equation}
We have therefore a very nice realization of holography, in which the dual field theory data is mapped on the $y=0$ two-plane of the gravity solution.

It is useful to describe the Fermi liquid using the variable $\rho(x^1,x^2)=1/2-z(x^1,x^2,0)$, which takes the value one inside the hall droplets, and zero outside. Then, it is tempting to go beyond the classical limit, and interpret the function $\rho$ as the probability density of the fermions. From the point of view of the SYM theory, as long as $\rho\in[0,1]$, one has a well-defined semi-classical state of the field theory yielding this probability density. 
Due to the noncommutativity of the phase space variables, the variation of $\rho
$ must be slow on the scale of $\sqrt\hbar$.
A $\rho>1$ would correspond to a violation of Pauli's exclusion principle, while negative density clearly has no meaning. However, on the gravitional side of the correspondence, {\em any} real function $\rho$ gives a one half BPS solution of the supergravity equations of motion, in general with some singularities.
As observed in \cite{CKS}, whenever the above physical conditions on $\rho$ are violated, the dual spacetime has closed timelike curves. This suggests that {\it the requirement of unitarity of the dual CFT implements the chronology protection for superstring theory on asymptotically \AdSxS spacetimes}.

For simplicity, we shall illustrate this behaviour in the case of the superstar solution, which is the simplest family of half BPS solutions of type IIB supergravity on \AdSxS{}, given by the metric
\begin{eqnarray}
ds^2=-\frac1{\sqrt{\Delta}}\left(\cos^2\theta+\frac{r^2}{L^2}\Delta\right)dt^2+
\frac{L^2H}{\sqrt\Delta}\sin^2\theta\,d\phi^2
+\frac{2L}{\sqrt\Delta}\sin^2\theta\,dt\,d\phi\nonumber\\
+\sqrt\Delta\left(\frac{dr^2}{f}+r^2\,d\Omega_3^2\right)
+L^2\sqrt\Delta\,d\theta^2+\frac{L^2}{\sqrt\Delta}\cos^2\theta\,d\tilde\Omega_3^2\,,\qquad\qquad
\label{superstar}\end{eqnarray}
with $H=1+Q/r^2$, $f=1+Hr^2/L^2$ and $\Delta=\sin^2\theta+H\cos^2\theta$. The ${\mathcal S}^5$ factor of the metric in these coordinates is given by
$d\theta^2+\sin^2\theta\,d\phi^2+\cos^2\theta\,d\tilde\Omega_3^2$.
This solution is obtained by oxidation to ten dimensions \cite{Cvetic:1999xp} of the extremal limit of the black hole solution with one R-charge of the $STU$ model \cite{Behrndt:1998ns,Behrndt:1998jd}. Again, there is a five-form field strength turned on, which can be found in \cite{Cvetic:1999xp}. For $Q=0$ we recover standard \AdSxS.
The $Q>0$ case exhibits a naked singularity at the origin $r=0$ of AdS, where a condensate of giant gravitons growing in the five-sphere sits, and acts as a source for the supergravity fields \cite{Myers:2001aq}. This view has been confirmed in \cite{Lin:2004nb}, where the corresponding fermion distribution in the dual CFT was interpreted as a dilute gas of holes in the Fermi sea. In fact, using the coordinate transformation
\begin{equation}
y=Lr\cos\theta\,,\qquad R=L^2\sqrt{f}\sin\theta\,,\qquad t\mapsto Lt\,,
\end{equation}
one recovers the LLM metric (\ref{LLM}) with polar coordinates $(R,\phi)$ on the $(x^1,x^2)$ plane. With some algebra, one can then compute the corresponding function $z(R,\phi,y)$, that reads
\[
z=\frac{1}{2(1+Q/L^2)}\left[ {y^2+R^2-R_0^2\over \sqrt{(y^2+R^2+R_0^2)^2-4R^2R_0^2}}+{Q\over L^2}\right]\,,
\]
where $R_0^2=L^2(L^2+Q)$. Its $y=0$ limit yields the fermion distribution
\begin{equation}
	\rho(R) = \left\{\begin{array}{c@{\qquad}l}
		\displaystyle\frac1{1+Q/L^2} & R < R_0\\
			                         &	                \\
		0                            & R > R_0\,.
	\end{array}\right.
\label{superstar-density}\end{equation}
LLM showed that the vacuum \AdSxS{} is represented by a Fermi droplet of density $\rho=1$ and radius $L^2$. Its total area $\pi L^4$ consists therefore exactly of $N$ Fermi cells of area $\hbar=2\pi l_p^4$. By turning on the R-charge $Q$, the probability density spreads to a disk of radius
$L^2\sqrt{1+Q/L^2}$,
but with lower density, in such a way that the correct number of fermions $\frac{1}\hbar\int\!\rho\,da=N$ is recovered. This fact supports the interpretation of $\rho$ as a density distribution of fermions.
The fermion system represents a uniform gas of holes delocalized in the Fermi sea, and since such holes correspond to giant gravitons growing on the five-sphere
\cite{Berenstein:2004kk,Caldarelli:2004ig}, the superstar can be thought of as the backreaction on spacetime produced by a condensate of giant gravitons \cite{Myers:2001aq}.

\begin{figure}[htbp]
\centering
\includegraphics[width=.7\linewidth]{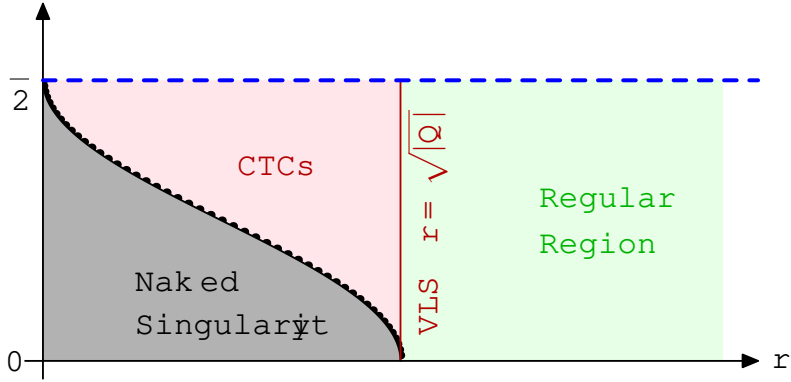}
\caption{Location of the naked singularity, the region with CTCs and the velocity of light surface (VLS) for the anti-superstar with $-L^2<Q<0$.}
\label{fig-superstar}\end{figure}

Let us suppose now that $-L^2<Q<0$ \footnote{The case with $Q<-L^2$ corresponds to putting sources away from the $(x^1, x^2)$-plane, and is therefore not covered by the LLM construction.}. The metric (\ref{superstar}) still represents a genuine solution of the supergravity theory. This spacetime is defined for
$r>\sqrt{|Q|}\cos\theta$, and has a naked singularity at $r=\sqrt{|Q|}\cos\theta$.
For $r>\sqrt{|Q|}$ it is perfectly well-behaved, but one checks that in-between,
for $\sqrt{|Q|}\cos\theta<r<\sqrt{|Q|}$, the coefficient $g_{\phi\phi}$ of the metric becomes negative, and the orbits of $\partial_\phi$ are CTCs (see figure~\ref{fig-superstar}).

Note that the metric remains lorentzian everywhere.
As in the superstar case, we can cast the metric in the LLM form to read off the source which produces this spacetime. The result is again (\ref{superstar-density}), but this time, since $Q<0$, the Fermi sea has been squeezed to a smaller radius, resulting in a density $\rho>1$. This implies the presence of more than one fermion per Fermi cell, in plain violation of Pauli's exclusion principle. The anti-superstar has no dual CFT state. Taking the dual CFT description equivalent to type IIB superstring theory as the fundamental one,
we see that the Pauli exclusion principle automatically rules out the solutions of the form (\ref{superstar}) with causal anomalies.

This mechanism works in fact in much more generality, and the chronology protection conjecture has been proven for general fermionic distributions with circular symmetry. Using continuity arguments, it was shown in \cite{CKS} that supergravity solutions resulting from non-physical distributions $\rho$ (i.e. which do not define a good probability density) have indeed CTCs. Subsequently, Milanesi and O'Loughlin proved the converse by a detailed analysis of the geometries, i.e. that solutions generated by well-behaved fermion densities are free of CTCs \cite{MO}.

To summarize, in the half BPS sector a spacetime suffers from CTCs if and only if the boundary condition $\rho(R)$ for the supergravity solution violates somewhere the bound $0\leq\rho(R)\leq1$. If we interpret $\rho$ as the probability density of fermions in the phase space, this bound corresponds to a unitarity bound of the dual field theory; $\rho>1$ would violate Pauli's exclusion principle, while negative $\rho$ would imply negative norm states. Therefore, configurations violating this bound do not correspond to any state in the Hilbert space of the SYM theory, which gives a fundamental description of superstring theory on \AdSxS, and the corresponding causality-violating backgrounds are forbidden. This mechanism implements in a simple form the chronology protection conjecture for a subsector of type IIB superstring theory in asymptotic \AdSxS{} backgrounds, and suggests a deep link between causality in gravitational theories and unitarity of the dual field theory.

Such a link also emerged in the analysis of the BMPV black hole \cite{Breckenridge:1996is,Herdeiro:2000ap}, for which a similar protection mechanism acts.
This black hole has CTCs, which are hidden by the event horizon in the underrotating regime, but develops naked CTCs if one increases the angular momentum and enters the overrotating regime.
Using the microscopic description in terms of a two-dimensional conformal field theory, provided by string theory, Herdeiro showed that the causality bound corresponds to a unitarity bound in the CFT \cite{Herdeiro:2000ap}, and therefore that the overrotating spacetimes are not genuine solutions of string theory. This analysis has then been generalized by Boni and Silva, who extended the protection mechanism to general AdS$_3$ bubbling solutions \cite{BS}.
Moreover, this mechanism could apply to non-supersymmetric configurations, and it was proposed in \cite{Caldarelli:2001iq} to hold for non-extremal Kerr-AdS$_5$ black holes of type IIB supergravity. Such solutions have naked CTCs for equal angular momenta and sufficiently negative mass. On the other hand, unitary representations of the superconformal algebra put a lower bound on the mass. It would be interesting to check if the bounds agree in this non-supersymmetric case.

Finally, it would be interesting to understand how this chronology protection mechanism can be realized in other sectors of AdS/CFT.


\section*{Acknowledgments}
This work was partially supported by INFN, MURST and
by the European Commission program MRTN-CT-2004-005104.


\end{document}